\documentclass[useAMS,usenatbib]{mn2e}

\usepackage{graphicx,amssymb,bm}
\usepackage{subfig,xfrac}
\usepackage{float, Abbrev}
\usepackage[fleqn]{amsmath}  
\usepackage{color,url}
\usepackage{upgreek}

\newcommand{\be}{\begin{equation}}
\newcommand{\ee}{\end{equation}}
\newcommand{\ba}{\begin{eqnarray}}
\newcommand{\ea}{\end{eqnarray}}

\usepackage{orcidlink}

\def\simlt{\lower.5ex\hbox{$\; \buildrel < \over \sim \;$}}
\def\simgt{\lower.5ex\hbox{$\; \buildrel > \over \sim \;$}}

\usepackage{graphicx} 

\title[Weak lensing shear measurement for JWST]{Weak gravitational lensing measurements of Abell\,2744 using JWST and shear measurement algorithm {\tt pyRRG-JWST}}

\author[Harvey \& Massey]{David R.\ Harvey\,\orcidlink{0000-0002-6066-6707}$^{1}$\thanks{E-mail: david.harvey@epfl.ch} and
Richard Massey\,\orcidlink{0000-0002-6085-3780}$^{2}$,
\\
$^{1}$Laboratoire d’Astrophysique, \'Ecole Polytechnique F\'ed\'erale de Lausanne (EPFL), Observatoire de Sauverny, CH-1290 Versoix, Switzerland\\
$^{2}$Centre for Extragalactic Astronomy, Department of Physics, Durham University, South Road, Durham DH1 3LE, UK}

\begin{document}

\date{Accepted ---. Received ---; in original form \today.}

\pagerange{\pageref{firstpage}--\pageref{lastpage}} \pubyear{2013}

\maketitle

\label{firstpage}

\begin{abstract}
\noindent 
We update the publicly available weak lensing shear measurement algorithm {\tt pyRRG} for the James Webb Space Telescope, and apply it to UNCOVER DR1 imaging of galaxy cluster Abell~2744.
At short wavelengths ($<2.5\,\upmu$m), shear measurements are  consistent between independent observations through different JWST bandpasses, and calibrated within 1.5\% of those from the Hubble Space Telescope.
At longer wavelengths, shear is underestimated by $\sim$$5\%$, probably due to coarser pixellisation. 
We model the spatially varying Point Spread Function (PSF) using {\tt WebbPSF}, whose moments are within 0.05 of real stars near the centre of the mosaic, where there are sufficient stars to also generate an empirical model.
We measure shear from up to 162~galaxies~arcminute$^{-2}$ to derive a map of (dark plus baryonic) mass with 12~arcsecond (55~kpc) spatial resolution. All code, catalogues and maps are available from \url{https://github.com/davidharvey1986/pyRRG}.
\end{abstract}


\section{Introduction} \label{sec:intro}
Galaxy clusters are the largest known bound structures in our Universe. Containing massive red elliptical galaxies and hot ionised gas, but dominated by a much more massive cocoon of dark matter, they have proved to be ideal laboratories to study the properties of dark matter \citep[e.g.][]{persistentGGSL,Bahe2021,meneghetti_20,velDepCross,bulletclusterA, A520A,MS1054,A520B,Harvey15}. Although dark matter seems not to interact with photons, so cannot be seen directly, it can be detected via gravitational lensing, the deflection of light from distant sources \citep[see reviews][]{DMclustersreview,clusterWLrev}. Gravitational lensing also magnifies the images of galaxies behind the cluster, presenting an opportunity to study the distant Universe at high resolution \citep[e.g.][]{furtak21,bouwens22,atek23}.

Giant arcs produced by strong gravitational lensing were first properly resolved by the Hubble Space Telescope (HST). The James Webb Space Telescope (JWST) offers similarly transformative increases in imaging resolution and depth.
In particular, weak gravitational lensing is the coherent distortion to the images of distant galaxies, whose light passes through the cluster on adjacent lines of sight. It is possible to measure the distorted shapes of these galaxies, but the dominant source of noise is the intrinsic variety of their shapes. If these are random, and with the empirical observation that the dispersion of their ellipticity is $\sigma_\gamma\approx0.3$ \citep{COSMOSintdisp} we can achieve weak lensing signal to noise noise ratio $\sim$~1 along a typical line of sight near a cluster by averaging the shapes of $\sim50$ galaxies. The resolution with which dark matter can be mapped is thus limited by the density of resolved distant galaxies. The resolution and depth of JWST imaging reveals a higher density of galaxies than ever before seen \citep[c.f.][]{Finner2023}. This should allow us to detect subtle changes in the the distribution and dynamics of dark matter predicted by alternative particle models of dark matter \citep[e.g.][]{SIDMSim,RobertsonBAHAMAS,SIDM_shapes_subhalo}. 

As an example of the power of JWST we shall measure weak gravitational lensing by one of the most massive known galaxy clusters, Abell 2744 (A2744, at RA $00$h $14$m $20$s, 
Dec $-30^\circ$ $23\arcmin$ $19\arcsec$, redshift $z=0.308$).
This is an ongoing merger between several components, each of mass $>10^{14}M_\odot$. It has been well studied, but not all those studies agree, even about the relative masses of components -- and the merger history that led to the observed configuration of galaxies, gas, and dark matter is complex \citep{A2744}. The cluster was first studied as one amongst a large sample of clusters, to investigate the observed discrepancy between X-ray masses and gravitational lensing masses  \citep{allen1998}. 
In the first dedicated analysis, \citet[hereafter M11]{A2744} used HST imaging to find strong lensing \citep{Zitrin2010} then combined it with measurements of weak lensing to derive a `non-parametric' (adaptive pixel grid) mass map.
\citet[hereafter M16]{A2744_medezinski} used imaging from the Subaru telescope to construct a `nonparametric' (pixellated) mass map using weak lensing but over a wider area. Abell 2744 was then selected as part of the Hubble Frontier Fields deep imaging survey, and a broad collaboration across the lensing community produced some of the highest resolution strong and weak lensing mass maps with Hubble \citep{HFF}. \citet[hereafter J16]{substructure_a2744} and \cite{Mahler_A2744} found significantly higher masses in all regions of the cluster, using the strong and weak lensing algorithm {\tt Lenstool} \citep{lenstool}. \cite{GRALE_A2744} again used both strong and weak lensing, and used a different non-parametric mass mapping technique, showing that the assumption that light traces mass (assumed by parametric methods) holds. The recent acquisition of JWST imaging has led to the publication of several more strong lensing analyses \citep[][and this paper]{bergamini23a,bergamini23b,Furtak23}. Here we shall attempt to reconcile the apparent discrepancies between groups of these independent analyses.

Most importantly, we release open source code {\tt pyRRG-JWST} to measure the weak lensing signal and reconstruct the distribution of mass in any future JWST NIRCam data. We use this opportunity to test for systematic or calibration biases in the method, and understand how uncertainty in JWST's Point Spread Function (PSF) affects measurements of shear. 


\section{Weak lensing theory}

Weak gravitational lensing is the apparent distortion of a spatially extended background source of light by foreground matter. 
Following the notation of \cite{NarayanBartelmann1996}, if the distribution of foreground matter is thin compared to the distance to the source, its three dimensional Newtonian potential, $\Phi$, can be considered in two dimensional projection 
\be
 \Psi = \frac{D_{\rm LS}}{D_{\rm LO}D_{\rm SO}}\frac{2}{c^2}\int\Phi(x,y,z) ~dz,
\ee
where $D$ is the angular diameter distance between the lens (L), source (S) and observer (O). 
Such a distribution of mass deflects passing rays of light by angle $\hat{\alpha}$, where the reduced deflection angle
\be
 \alpha = \frac{D_{\rm LS}}{D_{\rm SO}}\hat{\alpha}=\nabla \Psi.
 \ee

\if false
Rays of light are deflected by angle $\hat{\alpha}$ and appear to reach an observer from location $\theta$ instead of true location
\be
\beta = \theta - \frac{D_{\rm LS}}{D_{\rm S}}\hat{\alpha} = \theta - \alpha.
\ee
The reduced deflection angle, $\alpha$ 
\be
 \alpha = \nabla \Psi,
\ee
depends on the two dimensional gradient of the three dimensional Newtonian potential, $\Phi$, after it is projected along the line of sight,
\be
 \Psi = \frac{D_{\rm s}}{D_{\rm L}D_{\rm S}}\frac{2}{c^2}\int\Phi(D_{\rm L},\theta,z) dz.
\ee
Second derivatives of the potential are related to the projected mass density, $\kappa$,
\be
\kappa = \frac{\Sigma}{\Sigma_{\rm crit}} = \frac{1}{2}(\Psi_{11} + \Psi_{22}),
\ee
where $\Sigma$ is the projected surface density and $\Sigma_{\rm crit}$ is the normalisation based on the geometrical configuration of the lens,
\be
\Sigma_{\rm crit} = \frac{c^2}{4\pi G} \frac{D_{\rm so}} {D_{\rm ls}D_{\rm lo}},
\ee
and $D$ corresponds to the angular diameter distances separating the lens (l), the source (s) or the observer (o). 

Physically, the convergence corresponds to a scalar increase or decrease in the size and magnification of a distorted background source. In addition to this the source is sheared, represented by $\gamma$, a two component vector field,
\be
\gamma_1 = \frac{1}{2}(\Psi_{11} - \Psi_{22})~~~~{\rm and }~~~~ \gamma_2 = \Psi_{12} = \Psi_{21},
\ee 
corresponding to a stretch along the x-axis for $\gamma_1$ and $45^\circ$ for $\gamma_2$. We now have a relation between the observable distortion and the lensing potential. The underlying degeneracy between shear and convergence means that  one cannot be observed without the other, this is known as the reduced shear,
\be
g=\gamma / (1 - \kappa).
\ee

\fi

Resolved background galaxies appear distorted if light from one side is deflected more than light at the other side. This is created by nonzero second derivatives of the potential 
\be
\kappa = \frac{1}{2}(\Psi_{11} + \Psi_{22}), ~~~
\gamma_1 = \frac{1}{2}(\Psi_{11} - \Psi_{22}),~~{\rm and }~~ \gamma_2 = \Psi_{12} = \Psi_{21}, \label{eqn:kap_shear}
\ee 
where convergence $\kappa$ is an isotropic magnification, and shear $\gamma_1$ ($\gamma_2$) is an elongation in the East-West (Northwest-Southeast) direction. 


From measurements of shear $\gamma_i$ it is possible to calculate the convergence via either Bayesian inference to fit parametric models \citep[e.g.][]{lenstool} or directly via Fourier space \citep{KS93}. The Fourier space inversion exploits the fact that both shear and convergence are derivatives of the same lensing potential so, following equation~\eqref{eqn:kap_shear},
\be
\tilde{\gamma_i}=
\left[\begin{matrix}
(k_1^2-k_2^2)/k^2 \\
2k_{2}k_1/k^2
\end{matrix}\right]
\tilde{\kappa}, \label{eq:ks93}
\ee
where tildes denote Fourier transforms, $k$ is the wavenumber, and a complex field is obtained for convergence $\kappa=\kappa_E+i\kappa_B$, such that $\kappa_E$=$\kappa$ and $\kappa_B$ should be zero in the absence of systematic bias.

Finally, we calculate the projected surface mass density
\be
\Sigma = \left[ \frac{c^2}{4\pi G} \frac{D_{\rm SO}} {D_{\rm LS}D_{\rm LO}} \right] \kappa. \label{eq:sigma_kappa}
\ee
Notice how the prefactor depends only on the geometrical configuration of the lens and source. Throughout this paper, we shall assume a Planck cosmology \citep{planckParsFinal} to convert between $\kappa$ and mass.

\section{Data}

We analyse reduced and stacked NIRCam imaging from DR1 of the JWST UNCOVER survey\footnote{\url{https://jwst-uncover.github.io/DR1.html}}, a $\sim$29~arcmin$^2$ mosaic near RA $00^\mathrm{h}$:$20^\mathrm{m}$:$00^\mathrm{s}$, Dec $-30^\circ$:$22\arcmin$:$30\arcsec$ \citep[see Figure~2 of][]{JWST_UNCOVER}. We attempt to measure the weak lensing signal independently in all available bands (f115w, f150w, f200w, f277w, f356w and f444w) to understand the behaviour of each (we do not include f090w since it does not have the same coverage as other bands). 
Parameters of the data relevant to weak lensing are summarised in table~\ref{tab:data}. 

\begin{table}
  \centering
\begin{tabular}{||c c c c c||} 
 \hline
Filter &  PixScale  & Kernel & PixFrac & ET (ks) \\
\hline
f115w & $0.02\arcsec$ & square & $0.75$  & $582.02$  \\
f150w & $0.02\arcsec$ & square & $0.75$  & $467.87$  \\
f200w & $0.02\arcsec$ & square & $0.75$  & $341.77$  \\
f277w & $0.04\arcsec$ & square & $0.75$  & $83.77$ \\
f356w & $0.04\arcsec$ & square & $0.75$  & $85.01$ \\
f410m & $0.04\arcsec$ & square & $0.75$  & $53.60$ \\
f444w & $0.04\arcsec$ & square & $0.75$  & $182.18$ \\
 \hline
\end{tabular}\caption{Overview of the JWST UNCOVER survey data \citep{JWST_UNCOVER}, including data reduction parameters relevant to weak lensing measurements: the filter, the final pixel scale, the `drizzle' kernel and `pixel fraction' used to resample and stack multiple exposures, and the total exposure time in kilo-seconds. \label{tab:data}}
\end{table}

\begin{table*}
  \centering
\begin{tabular}{||c c c c c c||} 
 \hline
Filter & Initial Cat & PSF Corrected & Size Cuts & Cluster Members  & Median Redshift \\
\hline
f115w & 1670 & 1166 &  187  & 138.6 & 1.36 \\
f150w & 1864 & 1349 &  202  & 151.4 & 1.49 \\
f200w & 1877 & 1374 &  216  & 161.9 & 1.56 \\
\hline
\end{tabular}\caption{Impact of various cuts in the catalogue to reach our final density of source galaxies for each filter given in units of galaxies per square arc-minute. The first column shows the size of the initial source catalogue, followed by the size after star-galaxy separation and PSF correction, followed by size and magnitude cuts, then the removal of cluster members, then the drop due to matching to the photo-z catalogues then the final redshift cuts. \label{tab:cuts}}
\end{table*}

\section{Shear measurement method}

To measure weak lensing shear from galaxies in the A2744 field, we adapt the publicly available code {\tt pyRRG} \citep{RRG,COSMOSintdisp,reconciling} to the specifics of JWST NIRCam data. This method is similar to the well-known KSB \citep{KSB} method, but corrects all moments of a galaxy's shape for convolution with the PSF before calculating an ellipticity, $e^\mathrm{obs}_i$, from a ratio of its Gaussian-weighted quadrupole moments. Specifically the two components of uncorrected ellipticity, $chi$ of a galaxy is defined by,
\be
\chi_1=\frac{J_{11}-J_{22}}{J_{11}+J_{22}}~~~~~~\chi_2=\frac{2J_{12}}{J_{11}+J_{22}},
\ee
where $J_{xx}$ is the quadrupole, normalised weighted image moment. 

Performing PSF correction first makes the calculation more stable in the presence of a diffraction-limited PSF whose extended wings mean that its moments converge slowly \citep[note that][propose an alternate solution to this problem]{fourier_quad}.
A local estimate of reduced shear can finally be obtained as $\tilde\gamma_i = e^\mathrm{obs}_i/GC$, where shear responsivity $G$ depends on the galaxy's higher order moments
\be
G=2 - \langle\chi^2\rangle-\frac{1}{2}\langle\lambda\rangle-\frac{1}{2}\langle\chi\cdot \mu\rangle
\ee
where $\langle\chi\cdot \mu\rangle=\chi_1\mu_1+\chi_2\mu_2$, $\lambda$ is a combination of the fourth order image moments with,
\be
\lambda=(J_{1111}+2J_{1122}+J_{2222})/(2d^2w^2),
\ee
where $d$ is the size of galaxy, $d=\sqrt{0.5(J_{11}+J_{22})}$, $w$ is a Gaussian weight function with a standard deviation of $d$, and the two components of $\mu$ are given by,
\be
\mu_1=(-J_{1111}+J_{2222})/(2d^2w^2),
\ee
\be
\mu_2=-2(J_{1112}+J_{1222})/(2d^2w^2).
\ee
Finally $C=0.86$ is a mean empirical calibration \citep{COSMOSintdisp}. \cite{High2007} demonstrates that pixellisation effects lead to different ideal values of $C_1$ for $\gamma_1$ and $C_2$ for $\gamma_2$. However, There is no unique direction of pixellisation in UNCOVER data, because the orientation or pixels is different in raw and stacked images. Moreover, since we are here interested only in nonlocal combinations of shear, the calibration averages over these two values: we have checked that neither our maps nor masses are changed withing statistical significance by any combination of $C_1$ and $C_2$ such that $(C_1+C_2)/2=0.86$.

\subsection{Object Detection}

We find objects in the stacked images using {\tt SExtractor} \citep{sextractor}, with ``hot'' and ``cold'' runs to improve deblending \citep{COSMOSintdisp}. We distinguish stars from galaxies using an interactive GUI to select loci in a space spanned by objects' peak surface brightness $\mu_\mathrm{max}$ and integrated magnitude. We then measure the second and fourth Gaussian-weighted moments of every star and galaxy, with the Gaussian centred such that the first order moment is zero \citep{RRG}.

\subsection{Correction for the Point Spread Function}

As we see above, we require both the second and fourth order moments in order to calculate the final shear. We thus must correct the galaxies' observed shape moments for convolution with for the JWST Point Spread Function (PSF), following \S5 of \cite{RRG}. This requires estimates of the second and fourth order image moments of the PSF, interpolated to the location of every galaxy. Specifically we correct the second order image moments, $J_{ij}$ of each galaxy via,
\be
J_{ij}=J'_{ij}-C'_{ijkl}P_{kl},
\ee
where $C'$ is convolution susceptibility tensor and is given by
\be
C_{ijkl}=\delta_{ik}\delta_{jl}-\frac{2}{w^2}J_{ki}\delta_{jl}+\frac{1}{2w^4}\left[J_{ijkl}-J_{ij}J_{kl}\right],
\ee
and $P$ is the unweighted PSF moments. Then the fourth order image moments via,
\be
J_{ijkl}=J'_{ijkl}-P_{ijkl}-6P_{ij}J'_{kl}+6P_{ij}P_{kl}.
\ee

We can calculate the second and fourth order moments of the PSF throughout the image in two ways. The first is to empirically measure the moments of the stars at different points in the image and then interpolate to the position of the galaxies we want to correct. The second is to use a model of the JWST PSF at exactly the position of each galaxy. Unfortunately, at galactic latitude $-81^\circ$, the UNCOVER images contain insufficient bright stars to do the first empirical method. We must therefore use a model and artificially plant stars in the JWST image, measure these moments and use these as the PSF moments to correct the galaxies with.

To do this we first create a JWST NIRCam image containing a dense grid of fake stars in each band, using the publicly available {\tt WebbPSF}\footnote{\url{https://webbpsf.readthedocs.io/en/latest/index.html}}. We measure the moments of these fake stars and fit a bivariate spline\footnote{\url{https://docs.scipy.org/doc/scipy/reference/generated/scipy.interpolate.RectBivariateSpline.html}} to interpolate to any point in the image. For the spline we use the default smoothing value recommended by {\tt scipy}: the length of the input vector, which roughly equates to the standard deviation of the values (i.e.\ the error in the estimates of the moments). We also use a degree=3 in each Cartesian direction. Then for each exposure that makes up the final drizzled image we calculate the moments for the PSF at the position of each galaxy, rotate it, and stack it across all exposures. This provides an model PSF at any point in the stacked image, accounting for rotations and discontinuities between exposures. 

\begin{figure*}
\begin{center}
\includegraphics[width=\textwidth]{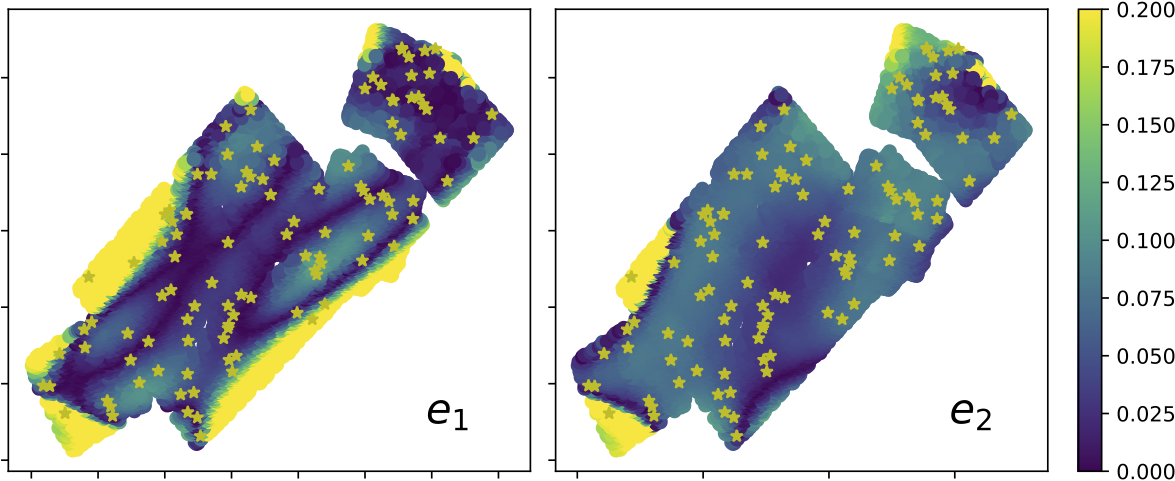}
\caption{The difference between models of the spatially-varying JWST PSF, calculated using {\tt WebbPSF} or empirically interpolated from real, bright stars in the stacked image mosaic (shown as yellow stars). The left (right) panel shows absolute differences in the first (second) component of ellipticity. The models match near the centre of the mosaic, where real stars are plentiful, but the empirical model diverges at the edges, due to the low density of stars at galactic latitude $-81^\circ$.}
\label{fig:psf}
\end{center}
\end{figure*}

To validate the PSF, we also use splines to interpolate the observed moments of stars in the stacked image. This avoids reliance on {\tt WebbPSF}, but does not account for discontinuities in the stacked image. Near the centre of the mosaic, whence there are stars in every direction, the difference between real stars and the {\tt WebbPSF} model is less than 0.05 in both components of ellipticity (Figure~\ref{fig:psf}). In the outskirts of the mosaic, the empirical interpolation struggles to converge. Nonetheless, we shall propagate these measurements through our full analysis, to quantify what difference they make to inferred cluster masses \citep[c.f.][]{Finner2023}, compared to the {\tt WebbPSF} model.


\subsection{Galaxy catalogue selection cuts}\label{sec:cuts}

\subsubsection{Size and brightness cuts}

The RRG shear estimator \citep[\S4.4 of][]{RRG}is a ratio of shape moments measured for each galaxy. The statistical weight applied to each galaxy sets a balance between reducing statistical noise by including more galaxies, versus growing systematic bias \citep{Refregier12} by including galaxies that are faint or small compared to the pixel scale. The optimum balance depends on the overall science goal, but for studies of individual clusters, we find a suitable compromise by a binary inclusion of galaxies with individual signal-to-noise ratio $>4.4$ \citep[as explored empirically by][]{COSMOSintdisp}. We also impose a radius cut of 4~pixels ($\sim$ 20\% larger than the PSF) and a minimum of 3\,FWHM distance between adjacent galaxies in crowded fields. This results in a shear catalgoue containing 145--168~galaxies~arcmin$^{-2}$, depending on the band (Table~\ref{tab:cuts}).



\subsubsection{Redshift cuts}\label{sec:zcuts}

Galaxies in (or in front of) the cluster will not be gravitationally lensed by it. Including them in any average shear measurement would spuriously dilute the shear signal and bias the inferred cluster mass.

\begin{figure}
\begin{center}
\includegraphics[width=0.5\textwidth]{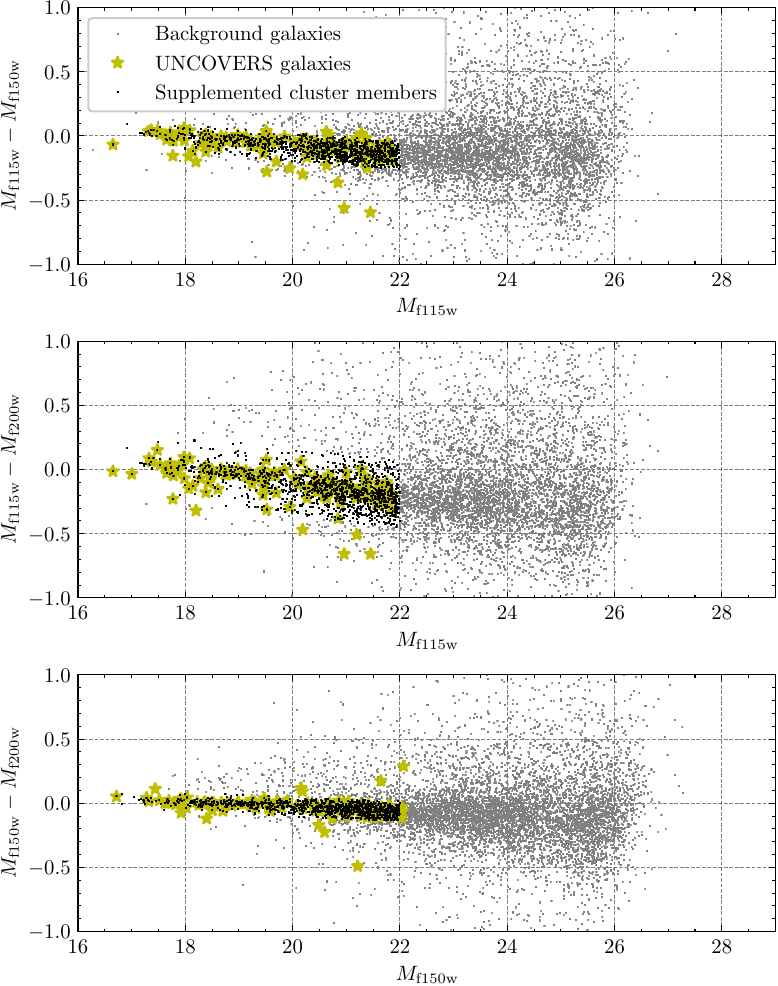}
\caption{
The shear catalogue excludes cluster member galaxies identified in combined JWST+HST photometry \citep[yellow stars;][]{JWST_UNCOVER}, but that covers only part of the JWST survey footprint. We use these to identify the cluster red sequence (bottom panel), and remove all galaxies in the red sequence throughout the survey.
}
\label{fig:clustermembers}
\end{center}
\end{figure}

To remove cluster member galaxies, we first note all galaxies in the \citep{JWST_UNCOVER} catalogue (other publicly-available catalogues are also available from \citealt{Furtak23} and \citealt{Weaver2024}). This is derived from HST imaging and does not completely cover the JWST field of view, so we supplement it by using the known cluster members to identify the red sequence in each combination of short JWST wavelengths (Figure~\ref{fig:clustermembers}). This sequence is tightest in the f150w and f200w bands since these bracket the Balmer break. We remove from our catalogue all galaxies within one sigma scatter of $|m_{\rm f150w}-m_{\rm f200w}|<0.08$ of that fitted line. 

To remove foreground galaxies, we also use UNCOVER photometry and photometric redshifts to identify all galaxies brighter than $m_{\rm f150w}<22$, or at redshift $z<0.350$. The latter accounts for photometric redshift uncertainty, and also ensures that all galaxies have high lensing efficiency. 

After cuts, the shear catalogues contain $\sim$150 galaxies/arcmin$^2$ in each band (see Table~\ref{tab:cuts} for the exact number in each band and the impact of each cut), at median photometric redshift $z\sim 1.72$ \citep{JWST_UNCOVER}. 

\subsection{Tests of shear calibration}

\subsubsection{Relative calibration between bands}

\begin{figure}
\begin{center}
\includegraphics[width=0.5\textwidth]{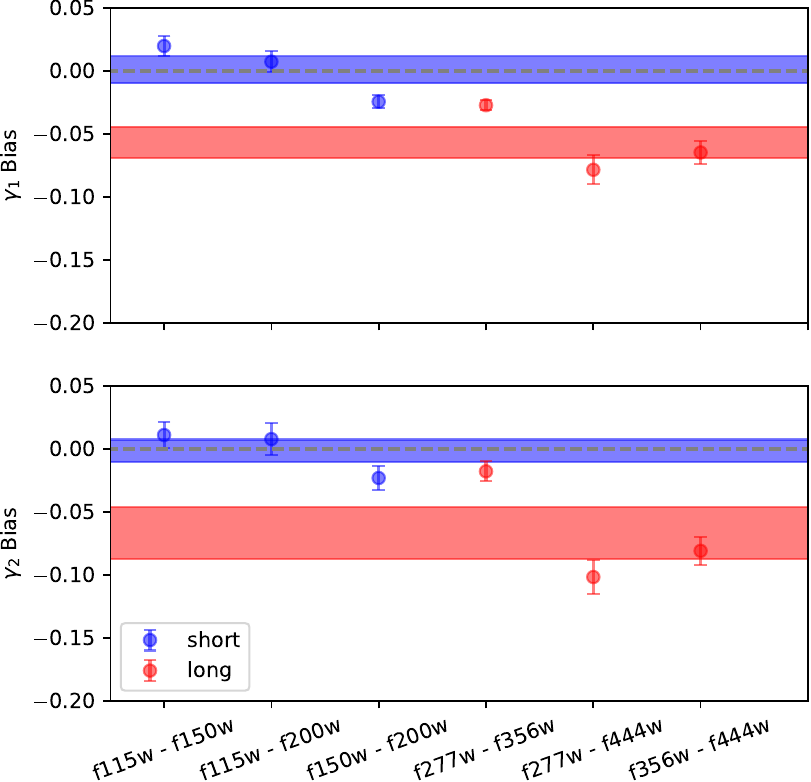}
\caption{Internal consistency check and estimate of multiplicative bias in our shape measurements. For a fixed sample of galaxies, we measure shear independently in all the UNCOVER imaging bands. The top (bottom) panel shows the relative calibration in $\gamma_1$ ($\gamma_2$) between bands, assuming a linear scaling: zero would indicate statistically identical calibration. Error bars show $1\sigma$ uncertainties. Coloured regions show the average bias for the various measurements at short (blue) and long (red) wavelengths. We find that the method has on average no significant bias in the short wavelengths and a $\sim5$\% bias in the longer wavelengths. }
\label{fig:shear_bias}
\end{center}
\end{figure}

We make independent measurements of galaxies' shear using every band of UNCOVER imaging. Since shear is independent of wavelength, shear estimators for a matched galaxy catalogue should be consistent with one another (they will not be identical because galaxies' intrinsic shapes $e^\mathrm{int}$ vary as a function of wavelength).

To compare shear measurements from two bands (say band A and band B), we use Orthogonal Distance Regression\footnote{\url{https://docs.scipy.org/doc/scipy/reference/odr.html}} to fit $\gamma^\mathrm{B}_i=(1+m_i)\gamma^\mathrm{A}_i + C_i$  \citep[c.f.][]{step1,step2}. Since imaging in different bands is drizzled to different pixel scales (see Table~\ref{tab:data}), we compare only shear measurements from similarly-drizzled bands, i.e.\ short wavelengths with short wavelengths and long wavelengths with long wavelengths. We find consistent behaviour in each regime, so also calculated the average biases between all pairs of bands at adjacent wavelengths (see Figure~\ref{fig:shear_bias}). At short wavelengths (where images have $0.02\arcsec$ pixels) there is no significant bias (between bands) with $m_1= 0.0\pm0.01$ and $m_2=0.001\pm0.009$, 68\% confidence limit. At long wavelengths (where images have $0.04\arcsec$ pixels), the typical bias is $m_1= -0.06 \pm0.01$ and $m_2=-0.07\pm0.02$, or less than 10\% at 68\% confidence limit. In all cases we find an additive bias consistent with zero.


\subsubsection{Relative calibration with respect to HST}

\begin{figure}
\begin{center}
\includegraphics[width=0.5\textwidth]{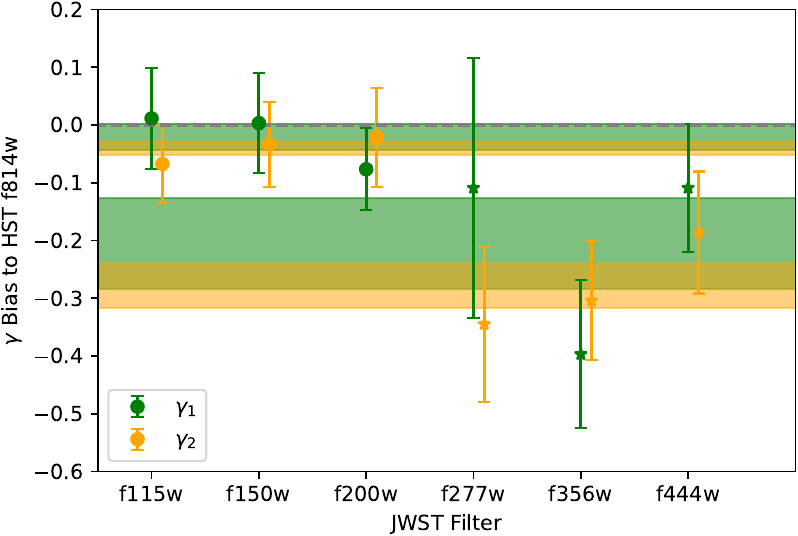}
\caption{Shear measurement bias in measurements with JWST, relative to a matched sample of galaxies from Hubble Space Telescope F814W imaging of the same cluster \citep[as measured by][]{Harvey15}. Green (orange) points show the mean relative bias in measurements of $\gamma_1$ ($\gamma_2$). At short JWST wavelengths, the mean shear measurement bias is within typical requirements for cluster lensing science goals. 
}
\label{fig:hst_bias}
\end{center}
\end{figure}

\begin{figure*}
\begin{center}
\includegraphics[width=0.49\textwidth]{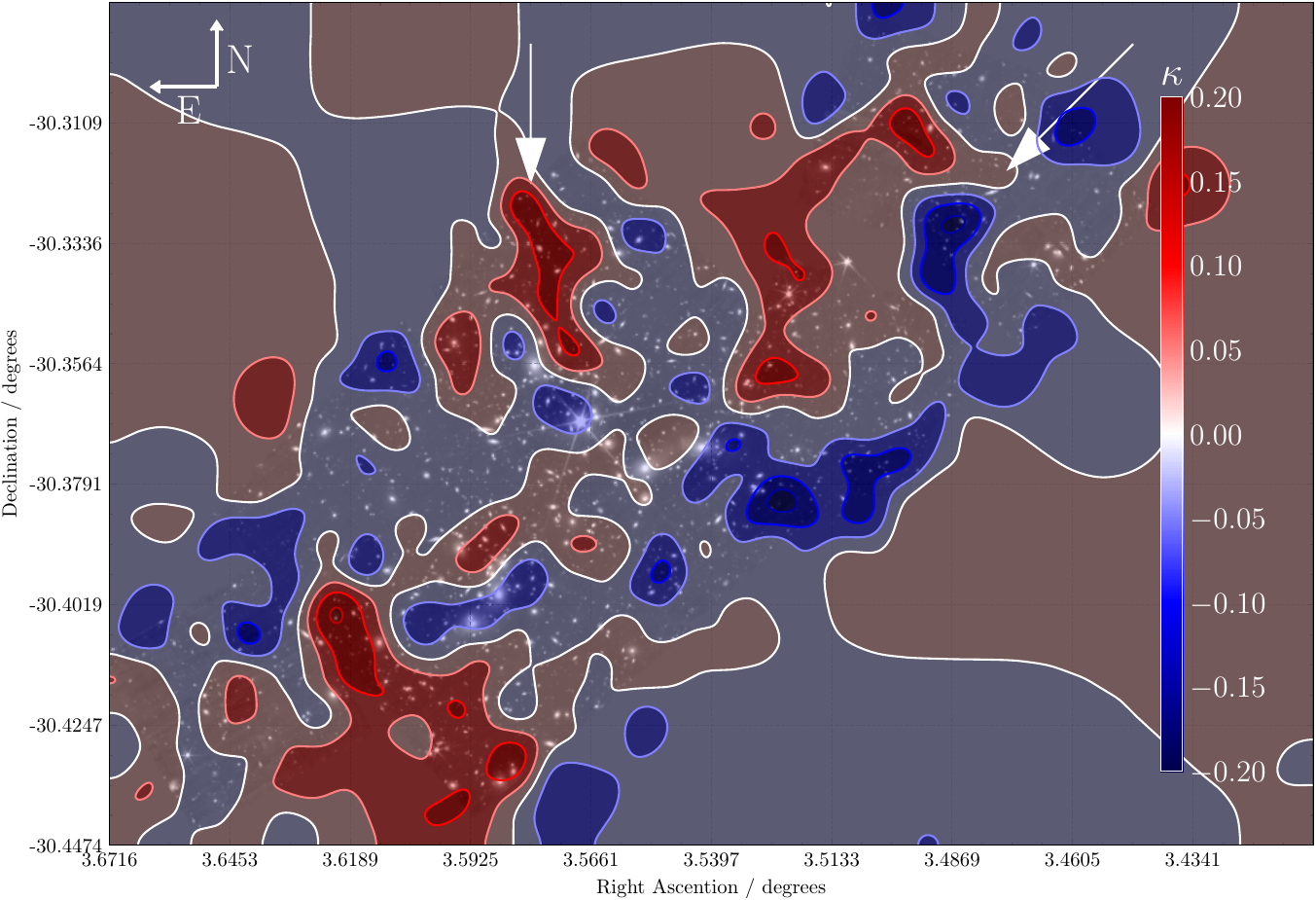}
\includegraphics[width=0.49\textwidth]{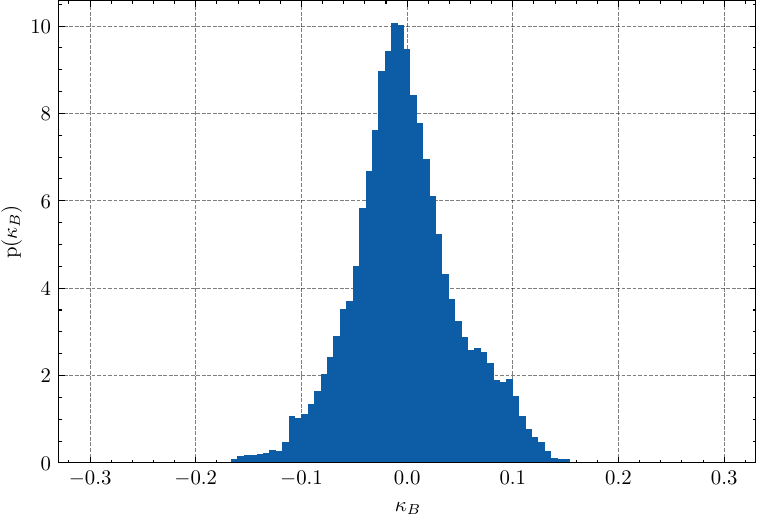}
\caption{ \emph{Left:} Galaxy cluster Abell~2744 weak lensing $B$-mode convergence map, which should be consistent with zero in the absence of systematic errors. \emph{Right:} A histogram of pixel values in the $B$-mode map shows a mean value consistent with zero, and standard deviation $\sigma_\kappa=0.05$, which should therefore also indicate the level of statistical noise in the $E$-mode map.}
\label{fig:final_map_bmode}
\end{center}
\end{figure*}

\begin{figure*}
\begin{center}
\includegraphics[width=\textwidth]{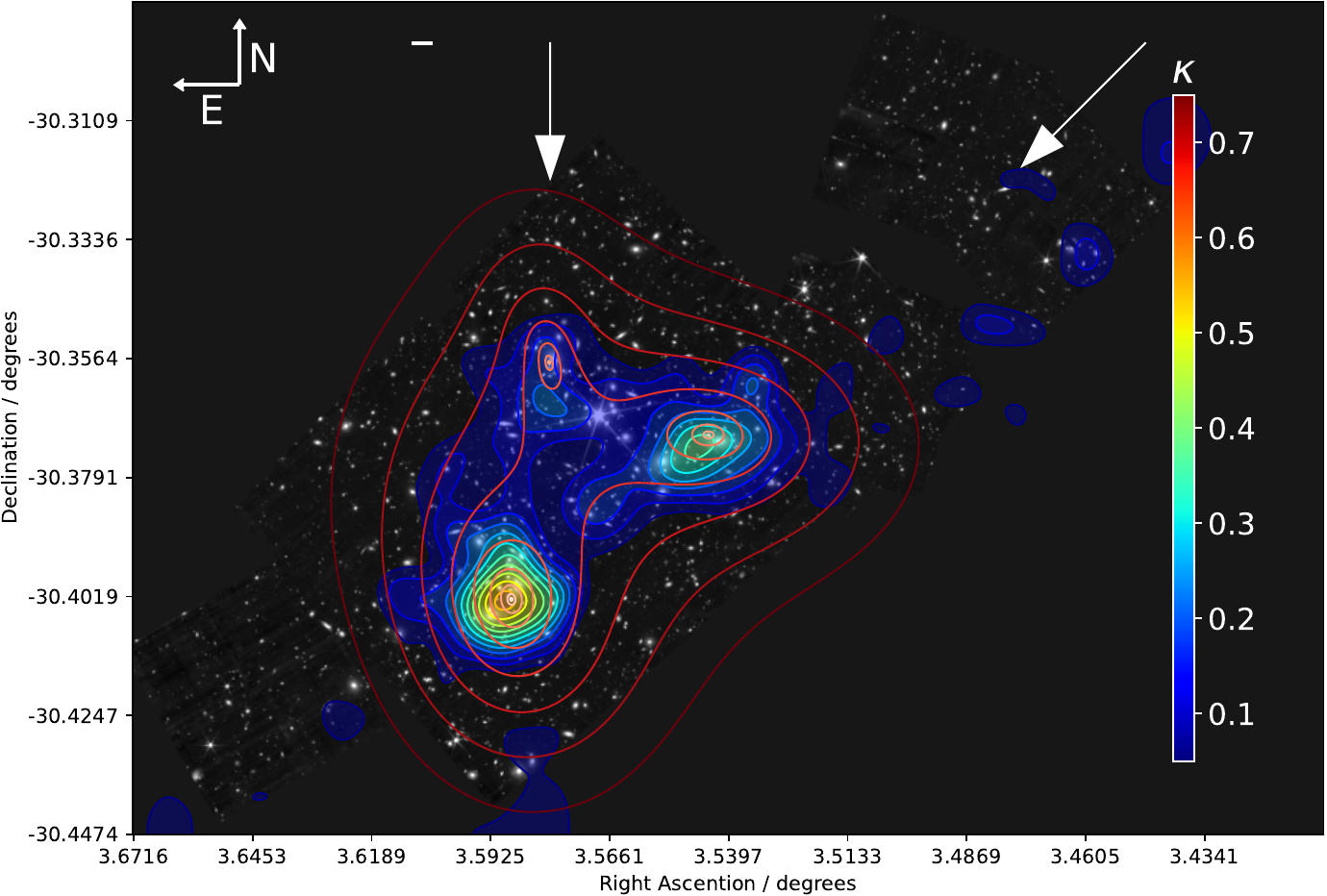}
\caption{Galaxy cluster Abell~2744 in JWST band f115w, plus non-parametric map of weak lensing $E$-mode convergence, which is proportional to total mass. The lensing analysis combines shear catalogues from imaging in f115w, f150w and f200w bands, all of which reach an average density of 150 source galaxies/arcmin$^2$. The lensing data are binned in 12.8~arcsec pixels and smoothed by a Gaussian filter of FWHM 17~arcsec, which optimises overall signal-to-noise. The first isodensity contour is at $\kappa=0.05$, with subsequent contours in steps of $\Delta\kappa=0.05$. We also show the best fit {\tt Lenstool} model in red contours. Arrows in the North and North-West indicate the direction of filaments identified by \citet{ISDC1}.}
\label{fig:finalmap}
\end{center}
\end{figure*}

The {\tt pyRRG} algorithm has been calibrated for HST observations in the f814w band using mock imaging with known shear \citep{COSMOSintdisp} and subsequently used in many analyses of real data \citep[e.g.][]{cosmos3dshear,A2744,Harvey15,substructure_a2744,Schrabback2018,Tam2020}. We match JWST galaxies with those detected in HST imaging by \cite{Harvey15}, and compute the same linear fit as above (see Figure~\ref{fig:hst_bias}). At shorter JWST wavelengths we find no significant bias $m_1$, but $m_2\approx3$\%. A limitation of this approach is that only the brighter and bigger galaxies detected by JWST are also detected by HST. Nonetheless, the measurements with these two different observatories is completely independent, so their consistency within these bounds is encouraging.

\subsubsection{Interpretation}

The {\tt pyRRG} algorithm appears to meet most cluster science requirements in JWST bands at $<2.5$\,$\upmu$m. The main cause of bias at longer wavelengths is probably the larger pixels. RRG treats pixellation as a component of the PSF, which is mathematically accurate to only first order, and creates biases when pixels are coarse \citep{High2007}. This effect is already accounted for in the empirical calibration of RRG at short wavelengths and small pixels \citep{COSMOSintdisp}.

\section{Results}

\subsection{Non-parametric mass map}


We average shear measurements from f115w, f150w, and f200w bands in a grid of 64$\times$44 square pixels, each $12.8\arcsec$ on a side and containing a median of 20 galaxies. We assume the shear in each pixel is  median redshift of every mean shear measurement is $z=1.72$.

There is a statistical excess of galaxies near cluster cores, probably caused by interloper cluster members in our catalogue, even after the cuts intended to remove them (see \S\ref{sec:zcuts}). To account for their dilution of the shear signal, we also pixellate a map of galaxy number density $n_{\rm gal}$ and smooth it with a Gaussian of width $\sigma=1$~pixel. We find that the median number of galaxies in each pixel is 11 and therefore in any pixel containing more than 11 galaxies, we multiply both components of $\tilde\gamma_i$ by $n_{\rm gal}/11$. 
Even the most populated pixel contains only 18 galaxies, and this procedure does not change the map within statistical precision.

We convert the pixellated shear map into a pixellated convergence map using equation~\ref{eq:ks93} \citep{KS93}. 
Since the orientation of the data means that the map contains a lot of empty pixels around it already, we do not need to pad with additional zeros during the Fourier transforms. We then account for the discrete nature of the sampling of this field by smoothing the convergence map by convolving with a Gaussian of standard deviation $12$~arcsec. We find that this amount of smoothing suitably balances precision (retaining the signal) and accuracy (suppressing the noise).
After smoothing, the imaginary component of reconstructed convergence, $\kappa_\mathrm{B}$, is consistent with zero, as it should be in the absence of systematic biases, and its pixel values have standard deviation $\sigma_{\kappa\mathrm{B}}=0.05$ (Figure~\ref{fig:final_map_bmode}). This empirical measurement of statistical precision incorporates all sources of noise in the shear catalogue, and should also apply to the real component of convergence, $\kappa_\mathrm{E}$.



\begin{table*}
  \centering
\begin{tabular}{||c c c c c c c||} 
 \hline
Clump & $x$ (arcsec) & $y$ (arcsec) & ellipticity $e$ & orientation $\theta$ [deg] & concentration $c$ &  $\log(M_{\rm 200} / (M_\odot)$ ) \\
\hline
Core & $-293^{+2}_{-2}$ & $-225^{+3}_{-2.30}$ & $0.17^{+0.06}_{-0.06}$ & $91^{+9}_{-10.37}$ & $3.4^{+0.7}_{-0.6}$ & $14.5^{+0.1}_{-0.1}$ \\
NW & $-160^{+5}_{-3}$ & $-118^{+2}_{-2.54}$ & $0.15^{+0.07}_{-0.06}$ & $20^{+17}_{-11.26}$ & $2.0^{+0.5}_{-0.4}$ & $14.4^{+0.1}_{-0.1}$ \\
N & $-261^{+6}_{-4}$ & $-70^{+7}_{-6}$ & $0.26^{+0.14}_{-0.13}$ & $119^{+20.84}_{-22}$ & $2.7^{+2.1}_{-1.2}$ & $13.8^{+0.3}_{-0.2}$ \\
 \hline
\end{tabular}\caption{Results of the {\tt Lenstool} fit to the combined filters, f115w, f150w and f200w using {\tt WebbPSF} to model the PSF. The offsets are within respect to the reference point $(3.556,-30.369)$.  \label{tab:lenstool_pars}}
\end{table*}

\begin{table*}
  \centering
\begin{tabular}{||c c c c c c c||} 
 \hline
Clump &  $M (<250$kpc$)$ $/10^{13}M_\odot$ & $M (<250$kpc$)$ $/10^{13}M_\odot$ &$M (<250$kpc$)$ $/10^{13}M_\odot$&$M (<250$kpc$)$ $/10^{13}M_\odot$  \\
           &This Work (WL)& \cite{substructure_a2744} (WL+SL)& \cite{A2744} (WL+SL) & \cite{A2744_medezinski} (WL) \\
\hline
Core  & $16.0_{-0.9}^{+0.6}$& $27.7 \pm 0.1$ & $22.4 \pm 5.5$ & $14.9 \pm 3.5$ \\
NW      &$10.8_{-1.0}^{+0.3}$&  $18.0\pm 1.0$ & $11.5 \pm 2.3$  &$ 7.6 \pm 3.5$ \\
N  & $6.5_{-0.9}^{+0.7}$ & $8.6 \pm 2.2$ &  --- & ---\\ 
 \hline
\end{tabular}\caption{Our results in the context of similar studies looking at A2744 with the probes (WL- Weak Lensing, SL- Strong Lensing) shown. We find that we under estimate the mass of the core with respect to J16 and M11 who both use strong lensing. However we are consistent with M16 who uses Subaru data. Our estimate of the North-Western and Northern halo are consistent with M11 however under-estimated with respect to J16. ).  \label{tab:m250}}
\end{table*}

\begin{figure}
\begin{center}
\includegraphics[width=0.5\textwidth]{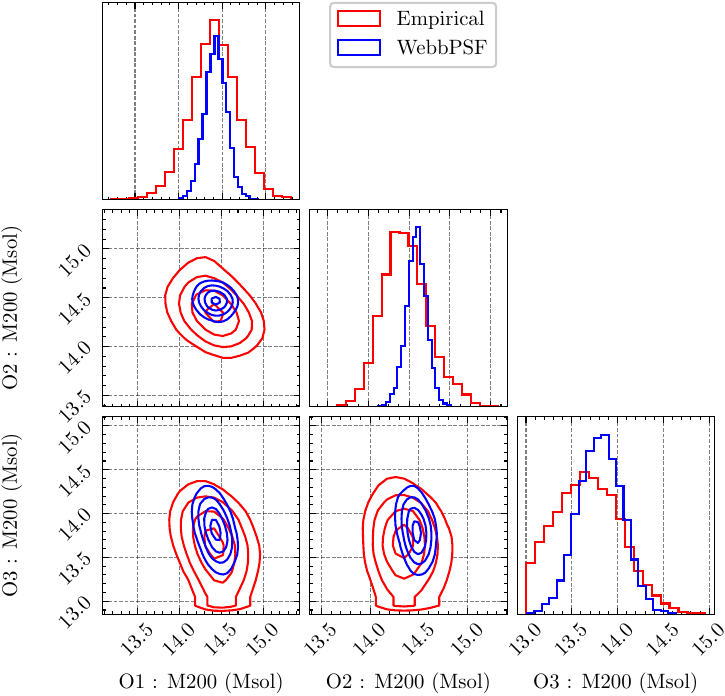}
\caption{Posterior probabilities for the $M_{200}$ mass of the three structures identified in the A2744 field. Blue contours show measurements from shears corrected for the PSF using {\tt WebbPSF}; red contours show measurements using an empirical PSF model interpolated from bright stars in the field. 
}
\label{fig:final_post}
\end{center}
\end{figure}

The $\kappa_\mathrm{E}$ convergence field contains three peaks, which match previous identification as Core, North-West and North substructures  \citep{A2744,substructure_a2744,Furtak23,bergamini23b}. Unfortunately the western halo found in previous studies lies just outside the UNCOVER survey footprint. The cluster member density and convergence trace each other extremely well, with filamentary extensions towards the North and North-West, corresponding to filaments identified by \citet{ISDC1}, and shown as arrows in figure~\ref{fig:finalmap}. 



\subsection{Parametric mass reconstruction}

To infer the mass of the three components using as much information as possible, we also fit a parametric model directly to the shear (and hence require no regularised grid) consisting of three \citep[NFW;][]{NFW} profiles. This has 18 free parameters: each component has unknown mass $M_{200}$, position ($x$, $y$), ellipticity ($e$, $\theta$), and concentration. We use relatively loose, flat priors on each position within $\pm25$~arcsec, the logarithm of halo mass between $10^{13}$--$10^{16}M_\odot$, concentration between $0$ and $20$, ellipticity between 0 and 1, and orientation of the major axis between $0^\circ$ and $180^\circ$.

To optimise the fit, we use the MCMC search inside {\tt Lenstool} \citep{lenstool}. Rather than binned shears, this takes measurements of every galaxy's shear and photometric redshift (which we extract from the UNCOVER catalogue; any galaxies without a photometric redshift we assume to be at $z=1.72$). Note that we do not attempt to correct for shear dilution by superfluous cluster member galaxies in the catalogue.

The best-fit model parameters have statistical uncertainty $\sim 2$\% (Table~\ref{tab:lenstool_pars}, with the posterior probability of cluster masses in Figure~\ref{fig:final_post}). Reassuringly, these observable quantities are robust within statistical uncertainty, regardless of the method used to model the PSF, or if shears without photometric redshifts are instead excluded, except that the uncertainities broaden. 

To compare our measurements to previous studies that state or for which it is possible to integrate the projected mass within a fixed radius, we similarly integrate our free-form mass map within $250$\,kpc of the best-fit centres (Table~\ref{tab:m250}). Our results are statistically consistent with M16, who used weak lensing measurements from Subaru imaging, and (in the North-Western and Northern structures) M11, who used weak and strong lensing measurements from HST. However, we measure a lower mass for the core than M11, where there is a high density of strong lensing constraints, and lower masses for all three structures than J16, who use only strong lensing. 


\section{Conclusions}

We use 6-band JWST-NIRCam imaging from UNCOVER data release 1 to measure weak gravitational lensing by the merging cluster A2744. The unmatched resolution and depth of JWST imaging achieve unprecedented resolution in the mass map, from shear measurements of $170$~galaxies/arcmin$^{2}$. However, consistent with previous studies, we identify three main mass components. The growing number of such independent analyses indicate that measurements from different telescopes are more consistent with each other than measurements using different (strong versus weak lensing) techniques. Our mass estimates are consistent with the weak lensing-only measurements of \cite{A2744_medezinski}, but are lower than the combined strong and weak lensing measurements of \cite{A2744} and \cite{substructure_a2744}. A similar dichotomy been seen before in other clusters. Identifying the cause will require further work, but in this case may also be exacerbated by uncertainty and bias in the photometric redshifts of very faint JWST galaxies.

For this kind of individual cluster lensing analysis, we find that the {\tt pyRRG} shear measurement method is sufficiently accurate to meet most science requirements. Measurements in bands at wavelengths $<2.5$\,$\upmu$m are consistent with each other and with measurements from HST: both comparisons indicate multiplicative shear measurement bias better than 2\%. The {\tt WebbPSF} model produces arbitrarily dense grids of stars throughout a mosaiced image, whose ellipticities agree with real stars within $\Delta e<0.05$. However, {\tt pyRRG} shear measurements in bands at longer wavelengths are not reliable, with multiplicative biases up to $\sim10\%$ between bands and $>30\%$ with respect to HST. This is likely because of the $0.04\arcsec$ pixels, which are larger than the $0.02\arcsec$ pixels at used at shorter wavelengths.

We publicly release all shear measurement code and catalogues at \url{https://github.com/davidharvey1986/pyRRG}, for any future work using weak lensing with JWST.

\section*{ Data Availability }
All code and data is publicly available at \url{https://github.com/davidharvey1986/pyRRG}
\section*{Acknowledgements}
This work was supported by the Swiss State Secretariat for Education, Research and Innovation (SERI) under contract number 521107294. 

\bibliographystyle{mn2e}
\bibliography{bibliography}

\begin{thebibliography}{50}
\expandafter\ifx\csname natexlab\endcsname\relax\def\natexlab#1{#1}\fi

\bibitem[{{Allen}(1998)}]{allen1998}
{Allen} S.~W., 1998, \mnras, 296, 392

\bibitem[{{Atek} {et~al}\mbox{.}(2023){Atek}, {Shuntov}, {Furtak}, {Richard}, {Kneib}, {Mahler}, {Zitrin}, {McCracken}, {Charlot}, {Chevallard}, \& {Chemerynska}}]{atek23}
{Atek} H. {et~al.}, 2023, \mnras, 519, 1201

\bibitem[{{Bah{\'e}}(2021)}]{Bahe2021}
{Bah{\'e}} Y.~M., 2021, \mnras, 505, 1458

\bibitem[{{Banerjee} {et~al}\mbox{.}(2020){Banerjee}, {Adhikari}, {Dalal}, {More}, \& {Kravtsov}}]{SIDM_shapes_subhalo}
{Banerjee} A., {Adhikari} S., {Dalal} N., {More} S., {Kravtsov} A., 2020, \jcap, 2020, 024

\bibitem[{{Bergamini} {et~al}\mbox{.}(2023{\natexlab{a}}){Bergamini}, {Acebron}, {Grillo}, {Rosati}, {Caminha}, {Mercurio}, {Vanzella}, {Angora}, {Brammer}, {Meneghetti}, \& {Nonino}}]{bergamini23a}
{Bergamini} P. {et~al.}, 2023{\natexlab{a}}, \aap, 670, A60

\bibitem[{{Bergamini} {et~al}\mbox{.}(2023{\natexlab{b}}){Bergamini}, {Acebron}, {Grillo}, {Rosati}, {Caminha}, {Mercurio}, {Vanzella}, {Mason}, {Treu}, {Angora}, {Brammer}, {Meneghetti}, {Nonino}, {Boyett}, {Brada{\v{c}}}, {Castellano}, {Fontana}, {Morishita}, {Paris}, {Prieto-Lyon}, {Roberts-Borsani}, {Roy}, {Santini}, {Vulcani}, {Wang}, \& {Yang}}]{bergamini23b}
{Bergamini} P. {et~al.}, 2023{\natexlab{b}}, \apj, 952, 84

\bibitem[{{Bertin} \& {Arnouts}(1996)}]{sextractor}
{Bertin} E., {Arnouts} S., 1996, \aaps, 117, 393

\bibitem[{{Bezanson} {et~al}\mbox{.}(2022){Bezanson}, {Labbe}, {Whitaker}, {Leja}, {Price}, {Franx}, {Brammer}, {Marchesini}, {Zitrin}, {Wang}, {Weaver}, {Furtak}, {Atek}, {Coe}, {Cutler}, {Dayal}, {van Dokkum}, {Feldmann}, {Forster Schreiber}, {Fujimoto}, {Geha}, {Glazebrook}, {de Graaff}, {Greene}, {Juneau}, {Kassin}, {Kriek}, {Khullar}, {Maseda}, {Mowla}, {Muzzin}, {Nanayakkara}, {Nelson}, {Oesch}, {Pacifici}, {Pan}, {Papovich}, {Setton}, {Shapley}, {Smit}, {Stefanon}, {Taylor}, \& {Williams}}]{JWST_UNCOVER}
{Bezanson} R. {et~al.}, 2022, arXiv e-prints, arXiv:2212.04026

\bibitem[{{Bouwens} {et~al}\mbox{.}(2022){Bouwens}, {Illingworth}, {Ellis}, {Oesch}, \& {Stefanon}}]{bouwens22}
{Bouwens} R.~J., {Illingworth} G., {Ellis} R.~S., {Oesch} P., {Stefanon} M., 2022, \apj, 940, 55

\bibitem[{{Clowe} {et~al}\mbox{.}(2004){Clowe}, {Gonzalez}, \& {Markevitch}}]{bulletclusterA}
{Clowe} D., {Gonzalez} A., {Markevitch} M., 2004, \apj, 604, 596

\bibitem[{{Clowe} {et~al}\mbox{.}(2012){Clowe}, {Markevitch}, {Brada{\v c}}, {Gonzalez}, {Chung}, {Massey}, \& {Zaritsky}}]{A520A}
{Clowe} D., {Markevitch} M., {Brada{\v c}} M., {Gonzalez} A.~H., {Chung} S.~M., {Massey} R., {Zaritsky} D., 2012, \apj, 758, 128

\bibitem[{{Eckert} {et~al}\mbox{.}(2015){Eckert}, {Jauzac}, {Shan}, {Kneib}, {Erben}, {Israel}, {Jullo}, {Klein}, {Massey}, {Richard}, \& {Tchernin}}]{ISDC1}
{Eckert} D. {et~al.}, 2015, \nat, 528, 105

\bibitem[{{Finner} {et~al}\mbox{.}(2023){Finner}, {Faisst}, {Chary}, \& {Jee}}]{Finner2023}
{Finner} K., {Faisst} A., {Chary} R.-R., {Jee} M.~J., 2023, \apj, 953, 102

\bibitem[{{Furtak} {et~al}\mbox{.}(2021){Furtak}, {Atek}, {Lehnert}, {Chevallard}, \& {Charlot}}]{furtak21}
{Furtak} L.~J., {Atek} H., {Lehnert} M.~D., {Chevallard} J., {Charlot} S., 2021, \mnras, 501, 1568

\bibitem[{{Furtak} {et~al}\mbox{.}(2023){Furtak}, {Zitrin}, {Weaver}, {Atek}, {Bezanson}, {Labb{\'e}}, {Whitaker}, {Leja}, {Price}, {Brammer}, {Wang}, {Marchesini}, {Pan}, {Dayal}, {van Dokkum}, {Feldmann}, {Fujimoto}, {Franx}, {Khullar}, {Nelson}, \& {Mowla}}]{Furtak23}
{Furtak} L.~J. {et~al.}, 2023, \mnras, 523, 4568

\bibitem[{Harvey {et~al}\mbox{.}(2015)Harvey, Massey, Kitching, Taylor, \& Tittley}]{Harvey15}
Harvey D., Massey R., Kitching T., Taylor A., Tittley E., 2015, Science, 347, 1462

\bibitem[{{Harvey} {et~al}\mbox{.}(2021){Harvey}, {Robertson}, {Tam}, {Jauzac}, {Massey}, {Rhodes}, \& {McCarthy}}]{reconciling}
{Harvey} D., {Robertson} A., {Tam} S.-I., {Jauzac} M., {Massey} R., {Rhodes} J., {McCarthy} I.~G., 2021, \mnras, 500, 2627

\bibitem[{{Heymans} {et~al}\mbox{.}(2006){Heymans}, {Van Waerbeke}, {Bacon}, {Berge}, {Bernstein}, {Bertin}, {Bridle}, {Brown}, {Clowe}, {Dahle}, {Erben}, {Gray}, {Hetterscheidt}, {Hoekstra}, {Hudelot}, {Jarvis}, {Kuijken}, {Margoniner}, {Massey}, {Mellier}, {Nakajima}, {Refregier}, {Rhodes}, {Schrabback}, \& {Wittman}}]{step1}
{Heymans} C. {et~al.}, 2006, \mnras, 368, 1323

\bibitem[{{High} {et~al}\mbox{.}(2007){High}, {Rhodes}, {Massey}, \& {Ellis}}]{High2007}
{High} F.~W., {Rhodes} J., {Massey} R., {Ellis} R., 2007, \pasp, 119, 1295

\bibitem[{{Jauzac} {et~al}\mbox{.}(2016){Jauzac}, {Eckert}, {Schwinn}, {Harvey}, {Baugh}, {Robertson}, {Bose}, {Massey}, {Owers}, {Ebeling}, {Shan}, {Jullo}, {Kneib}, {Richard}, {Atek}, {Cl{\'e}ment}, {Egami}, {Israel}, {Knowles}, {Limousin}, {Natarajan}, {Rexroth}, {Taylor}, \& {Tchernin}}]{substructure_a2744}
{Jauzac} M. {et~al.}, 2016, \mnras, 463, 3876

\bibitem[{{Jee} {et~al}\mbox{.}(2012){Jee}, {Mahdavi}, {Hoekstra}, {Babul}, {Dalcanton}, {Carroll}, \& {Capak}}]{A520B}
{Jee} M.~J., {Mahdavi} A., {Hoekstra} H., {Babul} A.~., {Dalcanton} J.~J., {Carroll} P., {Capak} P., 2012, \apj, 747, 96

\bibitem[{{Jee} {et~al}\mbox{.}(2005){Jee}, {White}, {Ford}, {Blakeslee}, {Illingworth}, {Coe}, \& {Tran}}]{MS1054}
{Jee} M.~J., {White} R.~L., {Ford} H.~C., {Blakeslee} J.~P., {Illingworth} G.~D., {Coe} D.~A., {Tran} K.-V.~H., 2005, \apj, 634, 813

\bibitem[{{Jullo} {et~al}\mbox{.}(2007){Jullo}, {Kneib}, {Limousin}, {El{\'{\i}}asd{\'o}ttir}, {Marshall}, \& {Verdugo}}]{lenstool}
{Jullo} E., {Kneib} J.-P., {Limousin} M., {El{\'{\i}}asd{\'o}ttir} {\'A}., {Marshall} P.~J., {Verdugo} T., 2007, New Journal of Physics, 9, 447

\bibitem[{{Kaiser} \& {Squires}(1993)}]{KS93}
{Kaiser} N., {Squires} G., 1993, \apj, 404, 441

\bibitem[{{Kaiser} {et~al}\mbox{.}(1995){Kaiser}, {Squires}, \& {Broadhurst}}]{KSB}
{Kaiser} N., {Squires} G., {Broadhurst} T., 1995, \apj, 449, 460

\bibitem[{{Leauthaud} {et~al}\mbox{.}(2007){Leauthaud}, {Massey}, {Kneib}, {Rhodes}, {Johnston}, {Capak}, {Heymans}, {Ellis}, {Koekemoer}, {Le F{\`e}vre}, {Mellier}, {R{\'e}fr{\'e}gier}, {Robin}, {Scoville}, {Tasca}, {Taylor}, \& {Van Waerbeke}}]{COSMOSintdisp}
{Leauthaud} A. {et~al.}, 2007, \apjs, 172, 219

\bibitem[{{Lotz} {et~al}\mbox{.}(2017){Lotz}, {Koekemoer}, {Coe}, {Grogin}, {Capak}, {Mack}, {Anderson}, {Avila}, {Barker}, {Borncamp}, {Brammer}, {Durbin}, {Gunning}, {Hilbert}, {Jenkner}, {Khandrika}, {Levay}, {Lucas}, {MacKenty}, {Ogaz}, {Porterfield}, {Reid}, {Robberto}, {Royle}, {Smith}, {Storrie-Lombardi}, {Sunnquist}, {Surace}, {Taylor}, {Williams}, {Bullock}, {Dickinson}, {Finkelstein}, {Natarajan}, {Richard}, {Robertson}, {Tumlinson}, {Zitrin}, {Flanagan}, {Sembach}, {Soifer}, \& {Mountain}}]{HFF}
{Lotz} J.~M. {et~al.}, 2017, \apj, 837, 97

\bibitem[{{Mahler} {et~al}\mbox{.}(2018){Mahler}, {Richard}, {Cl{\'e}ment}, {Lagattuta}, {Schmidt}, {Patr{\'\i}cio}, {Soucail}, {Bacon}, {Pello}, {Bouwens}, {Maseda}, {Martinez}, {Carollo}, {Inami}, {Leclercq}, \& {Wisotzki}}]{Mahler_A2744}
{Mahler} G. {et~al.}, 2018, \mnras, 473, 663

\bibitem[{{Massey} {et~al}\mbox{.}(2007{\natexlab{a}}){Massey}, {Heymans}, {Berg{\'e}}, {Bernstein}, {Bridle}, {Clowe}, {Dahle}, {Ellis}, {Erben}, {Hetterscheidt}, {High}, {Hirata}, {Hoekstra}, {Hudelot}, {Jarvis}, {Johnston}, {Kuijken}, {Margoniner}, {Mandelbaum}, {Mellier}, {Nakajima}, {Paulin-Henriksson}, {Peeples}, {Roat}, {Refregier}, {Rhodes}, {Schrabback}, {Schirmer}, {Seljak}, {Semboloni}, \& {van Waerbeke}}]{step2}
{Massey} R. {et~al.}, 2007{\natexlab{a}}, \mnras, 376, 13

\bibitem[{{Massey} {et~al}\mbox{.}(2010){Massey}, {Kitching}, \& {Richard}}]{DMclustersreview}
{Massey} R., {Kitching} T., {Richard} J., 2010, Reports on Progress in Physics, 73, 086901

\bibitem[{{Massey} {et~al}\mbox{.}(2007{\natexlab{b}}){Massey}, {Rhodes}, {Leauthaud}, {Capak}, {Ellis}, {Koekemoer}, {R{\'e}fr{\'e}gier}, {Scoville}, {Taylor}, {Albert}, {Berg{\'e}}, {Heymans}, {Johnston}, {Kneib}, {Mellier}, {Mobasher}, {Semboloni}, {Shopbell}, {Tasca}, \& {Van Waerbeke}}]{cosmos3dshear}
{Massey} R. {et~al.}, 2007{\natexlab{b}}, \apjs, 172, 239

\bibitem[{{Medezinski} {et~al}\mbox{.}(2016){Medezinski}, {Umetsu}, {Okabe}, {Nonino}, {Molnar}, {Massey}, {Dupke}, \& {Merten}}]{A2744_medezinski}
{Medezinski} E., {Umetsu} K., {Okabe} N., {Nonino} M., {Molnar} S., {Massey} R., {Dupke} R., {Merten} J., 2016, \apj, 817, 24

\bibitem[{{Meneghetti} {et~al}\mbox{.}(2023){Meneghetti}, {Cui}, {Rasia}, {Yepes}, {Acebron}, {Angora}, {Bergamini}, {Borgani}, {Calura}, {Despali}, {Giocoli}, {Granata}, {Grillo}, {Knebe}, {Macci{\`o}}, {Mercurio}, {Moscardini}, {Natarajan}, {Ragagnin}, {Rosati}, \& {Vanzella}}]{persistentGGSL}
{Meneghetti} M. {et~al.}, 2023, \aap, 678, L2

\bibitem[{{Meneghetti} {et~al}\mbox{.}(2020){Meneghetti}, {Davoli}, {Bergamini}, {Rosati}, {Natarajan}, {Giocoli}, {Caminha}, {Metcalf}, {Rasia}, {Borgani}, {Calura}, {Grillo}, {Mercurio}, \& {Vanzella}}]{meneghetti_20}
{Meneghetti} M. {et~al.}, 2020, Science, 369, 1347

\bibitem[{{Merten} {et~al}\mbox{.}(2011){Merten}, {Coe}, {Dupke}, {Massey}, {Zitrin}, {Cypriano}, {Okabe}, {Frye}, {Braglia}, {Jim{\'e}nez-Teja}, {Ben{\'{\i}}tez}, {Broadhurst}, {Rhodes}, {Meneghetti}, {Moustakas}, {Sodr{\'e}}, {Krick}, \& {Bregman}}]{A2744}
{Merten} J. {et~al.}, 2011, \mnras, 417, 333

\bibitem[{{Narayan} \& {Bartelmann}(1996)}]{NarayanBartelmann1996}
{Narayan} R., {Bartelmann} M., 1996, arXiv e-prints, astro

\bibitem[{{Navarro} {et~al}\mbox{.}(1997){Navarro}, {Frenk}, \& {White}}]{NFW}
{Navarro} J.~F., {Frenk} C.~S., {White} S.~D.~M., 1997, \apj, 490, 493

\bibitem[{{Peter} {et~al}\mbox{.}(2013){Peter}, {Rocha}, {Bullock}, \& {Kaplinghat}}]{SIDMSim}
{Peter} A.~H.~G., {Rocha} M., {Bullock} J.~S., {Kaplinghat} M., 2013, \mnras, 430, 105

\bibitem[{{Planck Collaboration} {et~al}\mbox{.}(2018){Planck Collaboration}, {Aghanim}, {Akrami}, {Ashdown}, {Aumont}, {Baccigalupi}, {Ballardini}, {Banday}, {Barreiro}, {Bartolo}, {Basak}, {Battye}, {Benabed}, {Bernard}, {Bersanelli}, {Bielewicz}, {Bock}, {Bond}, {Borrill}, {Bouchet}, {Boulanger}, {Bucher}, {Burigana}, {Butler}, {Calabrese}, {Cardoso}, {Carron}, {Challinor}, {Chiang}, {Chluba}, {Colombo}, {Combet}, {Contreras}, {Crill}, {Cuttaia}, {de Bernardis}, {de Zotti}, {Delabrouille}, {Delouis}, {Di Valentino}, {Diego}, {Dor{\'e}}, {Douspis}, {Ducout}, {Dupac}, {Dusini}, {Efstathiou}, {Elsner}, {En{\ss}lin}, {Eriksen}, {Fantaye}, {Farhang}, {Fergusson}, {Fernandez-Cobos}, {Finelli}, {Forastieri}, {Frailis}, {Franceschi}, {Frolov}, {Galeotta}, {Galli}, {Ganga}, {G{\'e}nova-Santos}, {Gerbino}, {Ghosh}, {Gonz{\'a}lez-Nuevo}, {G{\'o}rski}, {Gratton}, {Gruppuso}, {Gudmundsson}, {Hamann}, {Handley}, {Herranz}, {Hivon}, {Huang}, {Jaffe}, {Jones}, {Karakci}, {Keih{\"a}nen}, {Keskitalo}, {Kiiveri}, {Kim},
  {Kisner}, {Knox}, {Krachmalnicoff}, {Kunz}, {Kurki-Suonio}, {Lagache}, {Lamarre}, {Lasenby}, {Lattanzi}, {Lawrence}, {Le Jeune}, {Lemos}, {Lesgourgues}, {Levrier}, {Lewis}, {Liguori}, {Lilje}, {Lilley}, {Lindholm}, {L{\'o}pez-Caniego}, {Lubin}, {Ma}, {Mac{\'{\i}}as-P{\'e}rez}, {Maggio}, {Maino}, {Mandolesi}, {Mangilli}, {Marcos-Caballero}, {Maris}, {Martin}, {Martinelli}, {Mart{\'{\i}}nez-Gonz{\'a}lez}, {Matarrese}, {Mauri}, {McEwen}, {Meinhold}, {Melchiorri}, {Mennella}, {Migliaccio}, {Millea}, {Mitra}, {Miville-Dech{\^e}nes}, {Molinari}, {Montier}, {Morgante}, {Moss}, {Natoli}, {N{\o}rgaard-Nielsen}, {Pagano}, {Paoletti}, {Partridge}, {Patanchon}, {Peiris}, {Perrotta}, {Pettorino}, {Piacentini}, {Polastri}, {Polenta}, {Puget}, {Rachen}, {Reinecke}, {Remazeilles}, {Renzi}, {Rocha}, {Rosset}, {Roudier}, {Rubi{\~n}o-Mart{\'{\i}}n}, {Ruiz-Granados}, {Salvati}, {Sandri}, {Savelainen}, {Scott}, {Shellard}, {Sirignano}, {Sirri}, {Spencer}, {Sunyaev}, {Suur-Uski}, {Tauber}, {Tavagnacco}, {Tenti}, {Toffolatti},
  {Tomasi}, {Trombetti}, {Valenziano}, {Valiviita}, {Van Tent}, {Vibert}, {Vielva}, {Villa}, {Vittorio}, {Wandelt}, {Wehus}, {White}, {White}, {Zacchei}, \& {Zonca}}]{planckParsFinal}
{Planck Collaboration} {et~al.}, 2018, arXiv e-prints

\bibitem[{{Refregier} {et~al}\mbox{.}(2012){Refregier}, {Kacprzak}, {Amara}, {Bridle}, \& {Rowe}}]{Refregier12}
{Refregier} A., {Kacprzak} T., {Amara} A., {Bridle} S., {Rowe} B., 2012, \mnras, 425, 1951

\bibitem[{{Rhodes} {et~al}\mbox{.}(2000){Rhodes}, {Refregier}, \& {Groth}}]{RRG}
{Rhodes} J., {Refregier} A., {Groth} E.~J., 2000, \apj, 536, 79

\bibitem[{{Robertson} {et~al}\mbox{.}(2019){Robertson}, Harvey, {Massey}, {Eke}, {McCarthy}, {Jauzac}, {Li}, \& {Schaye}}]{RobertsonBAHAMAS}
{Robertson} A., Harvey D., {Massey} R., {Eke} V., {McCarthy} I.~G., {Jauzac} M., {Li} B., {Schaye} J., 2019, \mnras, 488, 3646

\bibitem[{{Sagunski} {et~al}\mbox{.}(2021){Sagunski}, {Gad-Nasr}, {Colquhoun}, {Robertson}, \& {Tulin}}]{velDepCross}
{Sagunski} L., {Gad-Nasr} S., {Colquhoun} B., {Robertson} A., {Tulin} S., 2021, \jcap, 2021, 024

\bibitem[{{Schrabback} {et~al}\mbox{.}(2018){Schrabback}, {Applegate}, {Dietrich}, {Hoekstra}, {Bocquet}, {Gonzalez}, {von der Linden}, {McDonald}, {Morrison}, {Raihan}, {Allen}, {Bayliss}, {Benson}, {Bleem}, {Chiu}, {Desai}, {Foley}, {de Haan}, {High}, {Hilbert}, {Mantz}, {Massey}, {Mohr}, {Reichardt}, {Saro}, {Simon}, {Stern}, {Stubbs}, \& {Zenteno}}]{Schrabback2018}
{Schrabback} T. {et~al.}, 2018, \mnras, 474, 2635

\bibitem[{{Sebesta} {et~al}\mbox{.}(2019){Sebesta}, {Williams}, {Liesenborgs}, {Medezinski}, \& {Okabe}}]{GRALE_A2744}
{Sebesta} K., {Williams} L. L.~R., {Liesenborgs} J., {Medezinski} E., {Okabe} N., 2019, \mnras, 488, 3251

\bibitem[{{Tam} {et~al}\mbox{.}(2020){Tam}, {Jauzac}, {Massey}, {Harvey}, {Eckert}, {Ebeling}, {Ellis}, {Ghirardini}, {Klein}, {Kneib}, {Lagattuta}, {Natarajan}, {Robertson}, \& {Smith}}]{Tam2020}
{Tam} S.-I. {et~al.}, 2020, \mnras, 496, 4032

\bibitem[{{Umetsu}(2020)}]{clusterWLrev}
{Umetsu} K., 2020, \aapr, 28, 7

\bibitem[{{Weaver} {et~al}\mbox{.}(2024){Weaver}, {Cutler}, {Pan}, {Whitaker}, {Labb{\'e}}, {Price}, {Bezanson}, {Brammer}, {Marchesini}, {Leja}, {Wang}, {Furtak}, {Zitrin}, {Atek}, {Chemerynska}, {Coe}, {Dayal}, {van Dokkum}, {Feldmann}, {F{\"o}rster Schreiber}, {Franx}, {Fujimoto}, {Fudamoto}, {Glazebrook}, {de Graaff}, {Greene}, {Juneau}, {Kassin}, {Kriek}, {Khullar}, {Maseda}, {Mowla}, {Muzzin}, {Nanayakkara}, {Nelson}, {Oesch}, {Pacifici}, {Papovich}, {Setton}, {Shapley}, {Shipley}, {Smit}, {Stefanon}, {Taylor}, {Weibel}, \& {Williams}}]{Weaver2024}
{Weaver} J.~R. {et~al.}, 2024, \apjs, 270, 7

\bibitem[{{Zhang} {et~al}\mbox{.}(2015){Zhang}, {Luo}, \& {Foucaud}}]{fourier_quad}
{Zhang} J., {Luo} W., {Foucaud} S., 2015, \jcap, 2015, 024

\bibitem[{{Zitrin} {et~al}\mbox{.}(2010){Zitrin}, {Broadhurst}, {Umetsu}, {Rephaeli}, {Medezinski}, {Bradley}, {Jim{\'e}nez-Teja}, {Ben{\'\i}tez}, {Ford}, {Liesenborgs}, {de Rijcke}, {Dejonghe}, \& {Bekaert}}]{Zitrin2010}
{Zitrin} A. {et~al.}, 2010, \mnras, 408, 1916

\end{thebibliography}

\bsp
\label{lastpage}

\end{document}